\documentstyle [12pt,aasms4]{article}
\journalid{}{}
\articleid{}{}

\begin{document}
\def\lax    {\ifmmode{_<\atop^{\sim}}\else{${_<\atop^{\sim}}$}\fi}
\def\gax    {\ifmmode{_>\atop^{\sim}}\else{${_>\atop^{\sim}}$}\fi}
\def\gtorder{\mathrel{\raise.3ex\hbox{$>$}\mkern-14mu
             \lower0.6ex\hbox{$\sim$}}}
\def\ltorder{\mathrel{\raise.3ex\hbox{$<$}\mkern-14mu
             \lower0.6ex\hbox{$\sim$}}}

\title{THE EXTENDED POWER LAW 
 AS  INTRINSIC SIGNATURE FOR A BLACK HOLE}

\author{
Lev Titarchuk\altaffilmark{1,3} and Thomas Zannias\altaffilmark{1,2}}

\altaffiltext{1}{National Aeronautics and Space Administration, 
Goddard Space Flight Center
(NASA/GSFC), Greenbelt, MD 20771, USA; E-mail: titarchuk@lheavx.gsfc.nasa.gov}
\altaffiltext{2}{Institute de Fisica y Matematicas, Universidad 
Michocana, S.N.H. Edificio C-3 Morelia Mich, Mexico; E-mail:
zannias@ginette.ifm.umich.mx}
\altaffiltext{3}{George Mason University/Institute for 
Computational Sciences and Informatics}

\rm

\vspace{0.1in}

\begin{abstract}

We analyze the exact general relativistic exact integro-differential 
equation of radiative transfer
describing the interaction of low energy photons with a Maxwellian 
distribution of hot electrons in gravitational field of a 
Schwarzschild black. We  prove that due to Comptonization 
an initial arbitrary  spectrum of low energy photons unavoidably results in 
 spectra characterized by an extended power-law feature.
We examine the spectral index by using both analytical and numerical 
methods  for a variety of physical parameters as 
such  the plasma temperature and the mass accretion rate.
The presence of the event horizon as well as the behaviour of the 
null geodesics in its vicinity  largely determine the dependence 
of the spectral index  on the flow parameters.
We come to the conclusion that the bulk motion of a converging flow 
is more efficient in upscattering photons than thermal Comptonization
provided that the electron temperature in the flow is of order of 
a few keV or less.  In this case, the spectrum observed at infinity
consists of a soft component produced by those input  photons that  
escape after a few scatterings without any significant energy change 
and of hard component (described by  a power law) produced by 
the photons that underwent significant upscattering.  
The luminosity of the power-law component is relatively small compared 
to that of the soft component. For accretion into black hole
the spectral energy index of the power-law is always higher than one 
for plasma temperature of order 
of a few keV. This result suggests  that the bulk motion 
Comptonization might be responsible for  the power-law spectra seen in 
the black-hole X-ray sources.

\end{abstract}


\section{INTRODUCTION}

Do black holes interact with an accretion flow in such a way so a distinct 
observational signature entirely different from those associated with any 
other compact object exists? In other words can
the existence of a black hole be solely inferred from  the 
radiation observed at infinity ?

These are the crucial questions where theoreticians and observers are 
confronting nowadays. 
Even though there have been now accumulated the 
enormous observational evidences in favor of the existence of black holes 
still it is fair to say that their existence has not been established.
 Perhaps the proof their existence
would have been a much easier task if, for instance, 
an argument would have been advanced, which would:
% and theoretical work supporting the view 
%that black hole indeed is a part of our observable universe still we 
%believe there has not been in the literature advanced a convincing 
%arguments that:

a) single out the generic component (or components) of a black hole which 
is responsible for shaping up the unique observed feature associated with 
black holes

b) prove that indeed this generic component always results in the same 
observable feature independent of the environmental conditions that the 
black hole finds itself.

The luck of such an argument may be traced in the plethora of various 
accretion flows: accretion in a state of free fall, optically thin or 
optically thick,
accretion disks with or without relativistic corrections, shocked flows 
{\it etc.} Of course, this diversity of accretion models is highly
justified. On physical grounds one expects accretion flows describing 
 a solar-mass black hole accreting interstellar medium to be distinct from
those flows  describing accretion onto a black hole in a close binary system 
or from a supermassive black hole at 
the center of an AGN. Viewed from this angle, the detectability of a black
hole appears to be a rather frustrating issue since it is not clear 
a priori what type of the existing accretion models (if any) would  
describe a realistic accreting black hole.

In the present paper we shall show that may be not the case.
The distinct feature of black hole spacetime, as opposed to the 
spacetimes due to  other compact objects is the presence of the event 
horizon.
Near the horizon the strong gravitational field is expected 
to dominate the pressure forces and thus to drive the accreting material into 
a free fall. In contrast, for other compact objects the pressure 
forces are becoming dominant as their surface is approached, and thus free 
fall state is absent. We  argue that this difference is rather crucial, 
 resulting in an observational signature of a black hole.
Roughly, the origin of this signature is due to the inverse Comptonization
of low energy photons from fast moving electrons.
The presence of the low-energy photon component is expected to be generic 
due, for instance, to the disk structure near a black hole or to Bremstrahlung 
of the electron component from the corresponding proton one. 
The boosted photon component is characterized by a power law spectrum, 
and is entirely independent of the initial spectrum of the low-energy 
photons. The spectral index of the boosted photons is  determined 
by the mass accretion rate and the bulk motion plasma temperature only.  
A key ingredient
in proving our claim is the employment of the exact relativistic transfer 
describing the Compton scattering of the low-energy radiation field of the 
Maxwellian distribution of fast moving  electrons.

We will prove  that the power law  is always present as a part of the 
black hole spectrum in a  wide energy range. We 
investigate the particular case of a non-rotating Schwartzchild black hole 
powering  the accretion, leaving the case of a rotating black hole 
for the future analysis. 

The presence of the power law part in the upcomptonized spectra was 
rigorously proven by Titarchuk \& Lyubarskij 1995 (hereafter TL95). 
There it has been demonstrated that for the wide class of the electron 
distributions the power law is the solution of the full kinetic equation.
   
The importance of  Compton upscattering of low-frequency photons 
in an optically thick, converging flow has been understood
 for a long time.  
Blandford and Payne were the first to address this problem in a series of
 papers (Blandford \& Payne 1981 and  Payne \& Blandford 1981).

In the first paper  they derived the Fokker-Planck
radiative transfer equation which took into account
photon diffusion in space and energy, while in the second paper 
 they solved the Fokker-Planck radiative transfer 
equation in the case of the steady state, spherically symmetric,
super-critical accretion into a central black hole with the assumption
of a power-law flow velocity $v(r)\propto r^{-\beta}$ and neglecting 
thermal Comptonization.  For the inner boundary condition they assumed 
adiabatic compression of photons as $r \to 0$.  Thus, their flow extended from
$r=0$ to infinity.  They showed that all emergent spectra have a high-energy, 
power-law tail with index $\alpha=3/(2-\beta)$ (for free fall $\beta=1/2$
and $\alpha=2$), which is independent of the low-frequency
source distribution.  

Titarchuk, Mastichiadis \& Kylafis (1996) paper
(hereafter TMK96 and see also the extended version in TMK97) presents 
the  exact numerical and approximate 
analytical solutions of the problem  of spectral formation 
in a converging flow, taking into account the inner boundary condition,
the dynamical effects of the free fall, and the thermal motion of 
the electrons.

The inner boundary has been taken at {\it finite} radius with the spherical 
surface being considered as a fully absorptive. 

Titarchuk, Mastichiadis \& Kylafis (1996) have 
 used a variant of the  Fokker-Plank formalism 
where the inner boundary mimics a black-hole horizon; 
no relativistic effects (special or general) are taken into account.
 Thus their  results are instructively useful but 
 they are not directly comparable with the observations. 
{\it By using the numerical and analytical techniques they demonstrated 
that the extended power laws are 
present in the resulting spectra in addition to 
the blackbody like emission at lower energies.}

Zane, Turrola, Nobilli \& Erna (1996) presented a characteristic method 
and the code for 
solving the radiative transfer equation in differentially moving media in
a curved spacetime. Some applications, concerning hot and cold accretion
onto nonrotating black holes were discussed there. 
  
In our paper the full relativistic treatment is worked out 
in terms of the relativistic Boltzmann kinetic equation 
without recourse to the Fokker-Planck approximation in either configuration
and energy space. 

The  relativistic transport theory  was developed 
by Lindquist (1966). He presents the appropriate radiative transfer 
in the curve spacetime. For completeness of our paper we delineate 
some important points of that theory related with the application to the 
radiative transfer in the electron atmosphere. 
 
We demonstrate that the power-law spectra are produced 
when low-frequency photons are scattered in the Thomson regime 
(i.e. when the dimensionless photon energy $z^=E^{\prime}/m_ec^2$ measured
in the electron rest frame satisfies $z^{\prime}\ll 1$). 

The eigenfunction method for the Comptonization problem 
employed in this paper has been offered and developed by 
Titarchuk \& Lyubarskij 1995 (hereafter TL95).

Giesler \& Kirk 1997 have extended the TL95 treatment by accommodating 
an arbitrary anisotropy of the source function. 
Their results for the spectral index confirm those of TL95 
over a wide range of electron temperature and optical depth; the 
largest difference they found is 10\%, occurring at low optical
depth.  

%Our numerical calculations of the spectral index for the relativistic case 
%considered in this paper are also consistent with
%the results of Monte-Carlo simulations (Laurent \& Titarchuk 1997) within 
%a few percents. 
 
The spectral indices related with the eigenvalues of the problem 
are determined  as functions
of the optical depth of the accreting matter.  Thus, 
for the first time we are able to solve the 
full Comptonization problem in the presence of the bulk and thermal motions 
of electrons.

In \S~2 and Appendix A we will give the details of the derivation of 
the general relativistic 
radiative kinetic equation. Section 3 ( and some details in the Appendix B) 
presents the method. We describe the  method of separation of variables, 
and the reduction of the whole problem to the specific eigen-problem in 
the configuration space. We propose the numerical solution of this problem 
by using the iteration method (e.g. Sunyaev \& Titarchuk 1985) with 
integrating over characteristics (the photon trajectories in the presence 
of Schwarzschild black hole background). Finally, we summarize our work and 
draw conclusions in \S 4.

\section{THE MAIN EQUATION}

We begin with considering, background geometry,  described by the 
following line element:
$$
ds^2~=~-fdt^2~+~{{dr^2}\over{f}}~+~r^2d\Omega^2
\eqno(1)
$$
where, for the Schwarzschild black hole, $f=1-r_s/r$, $r_s=2GM/c^2$, 
and $t,~r,~\theta,~\varphi$ are the event coordinates with
  $d\Omega^2= d\theta^2 +\sin^2\theta d\varphi^2$.
$G$ is the gravitational constant and $M$ is the mass of a black hole.

In order to describe the photon radiation field we shall employ
the concept of the distribution function $N$. 
The distribution function $N(x, {\bf p})$ describes the number 
$dN$ of photons (photon world lines) which cross a certain spacelike 
volume element $dV$ at $x(t,r,\theta,\varphi)$, and whose $4-$ momenta 
${\bf p}$ lie within a corresponding 3-surface element $dP$ in momentum 
space. It is desirable to choose $dV$ and $dP$ to be coordinate-invariants. 
Thus $dN$ would be invariant as well and the same would be true 
of $N(x, {\bf p})$. 

In Appendix A we present the detailed derivation of  
the relativistic radiative transfer equation expressed through 
the distribution function $N(x, {\bf p})$ and the interaction density 
function $S(N)$ (see the definition of this function after Eq. A13). 
 
We will describe the electron component
by a local Maxwellian distribution (e.g. Landau \& Lifshitz 1980, 
and Pathria 1970 )
$$
F(r, P_e)dP~=~\displaystyle{\aleph^{-1}e^{\beta u_{\mu}P_e^{\mu}}}dP
\eqno(2)
$$
where $\aleph$ is the normalization constant.

One has to interpret $F(r,P_e)(-P^{a}_{e}n_{a})dPdV$ 
similar to the one implied by (A8), with the sole exception that considerations
are restricted on the electron phase space.
For our purpose an arbitrary electron momentum state $P_{e}$ ,
can be represented in the form:

$$
P_e=\left({1\over{\sqrt{1-V^2/c^2}}},~{{|V|\bf n_e}\over{\sqrt{1-V^2/c^2}}}
\right),
\eqno(3)
$$
where the ``the thermal three velocity'' ${\vec V}$ stands for a convenient
parameterization of the electron phase space.
We will take $\beta=m_ec^2/kT_e$, while $u_{\mu}$ stands 
for the hydrodynamical four velocity of the inflowing plasma
which may be represented relative to the local orthonormal frame  in the 
form
$$
u=(u^{o},u^{r})=\left({1\over{\sqrt{(1-v^{2})}}},~
-{{\bf v^{r}}\over{\sqrt{(1-v^{2})}}}
\right),
\eqno(4)
$$
with the negative sign in $u^{r}$
 takes into account the convergent 
nature of the fluid flow.
Note that, as a result of the hydrodynamic bulk motion, the
local Maxwellian distribution exhibits a coupling of the thermal
velocity ${\vec V}$ with the hydrodynamic bulk motion ${\vec v}$, and one gets 
$$
\beta u_{\mu}P_e^{\mu}=-{{m_ec^2}\over{kT_e}}
(1-v^2/c^2)^{-1/2}(1-V^2/c^2)^{-1/2}\left[1+\cos\theta{{Vv}\over{c^2}}
\right]
\eqno(5)
$$
We will discuss the coupling  effect in \S 4  and we 
will consider this issue in detail in our next publication. 

Within a $4-$volume $dW$ at the event $x$ there is a decrease in the original 
number of world lines due to absorption and scattering out of momentum 
range $dP$, given by (cf. the right hand side of Eq. A13)
$$
-\kappa(x,{\bf p})n(x)N(x,{\bf p})~dW~dP.
\eqno(6)
$$
Here $n(x)$ is the proper number density of the electrons interacting 
with  the photons, namely the number density of electrons as measured 
in their own local rest frame, and 
$\kappa(x,{\bf p})$ is the invariant absorption coefficient or invariant 
opacity. The  $\kappa-$  opacity is related to  
 the usual scattering cross-section $\sigma_s$ via expression 
 (see Lindquist 1966) 
$$
\kappa~=~E\cdot\sigma_s.
\eqno(7)
$$
On the other hand, there is increase due to pure scattering out of all 
other $4-$momentum ranges $dP^{\prime}$ into $dP$ given by 
$$
n(x)~dW~dP\int dP^{\prime}\kappa(x,{\bf p})\zeta(x;{\bf p}^{\prime}
\rightarrow{\bf p}) N(x,{\bf p}^{\prime}).
\eqno(8)
$$
Thus  the transition probability $\zeta(x;{\bf p}^{\prime})$ 
can be expressed in terms of 
 the differential cross section $d\sigma_s/(dEd\Omega)$ 
(see Lindquist 1966) as 
$$
\kappa(x,{\bf p}^{\prime})\zeta(x;{\bf p}^{\prime}
\to {\bf p}) = {{E^{\prime}}\over E}
{{d\sigma_s}\over{dEd\Omega}}.
\eqno(9)
$$
Taking into account only Compton scattering of photons off the background 
electrons, one may covariantly write the transfer equation (see Eq. A13) 
in the following form:
$$
p^{\alpha}{{DN}\over{dx^{\alpha}}}=
{\int}N(r,P')\kappa(x,{\bf p}^{\prime})\zeta(x;{\bf p}^{\prime}
\rightarrow{\bf p})d P^{\prime}
$$
$$-
N(r,P){\int}N(r,P')\kappa(x,{\bf p^{\prime}})\zeta(x;{\bf p}
\rightarrow{\bf p^{\prime}}) dP^{\prime}. 
\eqno(10)
$$
The first term in the 
right hand side describes the increase in the photon world lines 
over the infinitesimal phase space cell centered around $P$
 while the second term describes the processes 
of depletion. 

 Recall that the 
scattering cross-section of photon from an electron in the electron's 
rest frame is described by the Klein-Nishina formula
$$
\sigma(\nu\to\nu^{\prime},\xi)=
{{3}\over{16\pi}}n_e\sigma_T{{1+\xi^2}\over{[1+z(1-\xi)]^2}}
\times $$
$$\times\left\{1+{{z^2(1-\xi)^2}\over{(1+\xi^2)[1+z(1-\xi)]}}
\right\}\delta\left[\nu^{\prime}-{{\nu}\over{1+z(1-\xi)}}\right] 
\eqno(11)
$$
where $z= h\nu/m_ec^2$ is a dimensionless photon energy,
$\xi$ is the cosine of scattering angle, $\sigma_T$ is the Thomson 
cross$-$section  and that all quantities in the right hand of the above 
formula are 
computed in the rest frame of the electron one may explicitly write the 
transfer equation on the black hole background.

By rewriting (10) for the orthonormal frame of (1), [see Eq. (A25)]
we get  the following equation:
$$
\mu\sqrt{f} {{\partial N}\over{\partial r}}-
\nu\mu{{\partial \sqrt{f}}\over{\partial r}}{{\partial N}\over{\partial \nu}}
-(1-\mu^2)\left({{\partial \sqrt{f}}\over{\partial r}}-
{{\sqrt{f}}\over{r}}\right)\cdot {{\partial N}\over{\partial \mu}}
=
$$
$$
\int_0^{\infty}d\nu_1\int_{4\pi}d\Omega_1
\left[\left({{\nu_1}\over{\nu}}\right)^2
\sigma_s(\nu_1\to \nu, \xi)N(\nu_1,\mu_1,r) -
\sigma_s(\nu\to\nu_1, \xi)N(\nu,\mu,r)\right].
\eqno(12)
$$

The scattering kernel can be calculated by performing a Lorentz boost of
$\sigma_s$, multiplying it by $F(r,P_e)$ [see Eq.(2)] and 
integrating over $P_e$. Then, the scattering kernel is given by 

$$
\sigma_s(\nu\to \nu_1, \xi,\beta)={3\over16\pi}{{n_e\sigma_T}\over{\nu z}}
\int_0^{\pi} \sin\theta d\theta \int d^{3}{\bf v}{{F(r,P_e)}
\over{\gamma}}
$$
$$
\left\{1+\left[1-{{1-\xi}\over{\gamma^2 DD^{\prime}}}\right]^2+
{{z z^{\prime}(1-\xi)^2}\over{\gamma^2D D^{\prime}}}\right\}
\delta(\xi-1+\gamma D^{\prime}/z-\gamma D/z^{\prime}),
\eqno(13)
$$
where $D=1-\mu V$, $D_1=1-\mu^{\prime}V$, 
$\gamma=(1-V^2)^{-1/2}$ 
and $\xi={\bf\Omega^{\prime}}\cdot{\bf\Omega}$ is the cosine of 
scattering angle. In deriving the above equation we have chosen
$$
\aleph(\beta)=m_ec\int_0^{\pi} \int_0^c \exp(\beta u_{\mu}P_e^{\mu})
\gamma^5{{V^2}\over{c^2}}\sin{\theta}dv^2 d\theta
\eqno(14)
$$
so that the distribution of electrons is normalized by a fixed electron 
density $n$ as measured in the orthonormal frame associated with (1).

\section{THE METHOD OF SOLUTION}

\subsection[THE METHOD OF SOLUTION]{Separation variables}

As long as the ejected low energy photons satisfy $z^{\prime}=
h\nu_0/m_ec^2 \gamma \ll 1,$ the integration over incoming 
frequencies $\nu_0$ is trivially implemented provided
that the explicit function of $N(r,\nu_0,\nu, {\bf \Omega})$ is known.
Thus, we need  to describe the main properties of  Green 
function $N(r,\nu_0,\nu, {\bf \Omega})$ in a situation 
when the low-energy photons are injected into atmosphere with 
the bulk motion.

The power-law part of the spectrum (Sunyaev \& Titarchuk 1980, TL95)
 occurs at frequencies lower than that of Wien cut-off 
($E< E_e$, where $E_e$ is the average electron energy). 
In this regime  the energy change due to the recoil effect of the electron 
can be neglected in comparison with the Doppler shift of the photon. 
 Hence we can drop the third term in parenthesis and 
the term $\xi-1$ of the delta-function argument in the scattering kernel 
(13) transforming that into the classical Thomson scattering kernel
(cf. TL95 and Gieseler \& Kirk 1997).  

Now we  seek the 
 solution of the Boltzmann equation (12) with the aforementioned 
simplifications, in the form
$$
N(r,\nu,{\bf \Omega})=\nu^{-(3+\alpha)}J(r,\mu). 
\eqno(15)
$$
Then we can formally  get from (12) that
$$
\mu\sqrt{f}{{\partial J}\over{\partial r}}+
(\alpha+3) \mu {{\partial\sqrt{f} }\over{\partial r}}J-
(1-\mu^2)\left({{\partial\sqrt{f} }\over{\partial r}}-
{{\sqrt{f} }\over{r}}\right){{\partial J}\over{\partial\mu}}=
$$
$$
=n_e\sigma_T\left[-J+{1\over{4\pi}}\int_{-1}^{1}d\mu_1\int^{2\pi}_0
d\varphi R(\xi)J(\mu_1,\tau)\right].
\eqno(16)
$$
Here the phase function $R(\xi)$ 
$$
R(\xi)={3\over4} \int_0^{\pi}\sin\theta d\theta 
\int d^3{\bf v}{{F(r, P_e)}\over{\gamma^2}}
\left({{D_1}\over{D}}\right)^{\alpha+2}{{1}\over{D_1}}
[1+(\xi^{\prime})^2],
\eqno(17)
$$
where $\xi^{\prime}$ is the cosine of scattering angle between photon 
incoming and outgoing directions in the electron rest frame.
 
The reduced integro-differential equation is two dimensional and it can be 
treated and solved much easer than the original equation (12).
The whole problem is reduced to the eigenvalue problem for equation (16). 
We can not claim that the kinetic equation allows a power-law solution (15)
unless first $\alpha$ is found and  $J(r,\mu)$ is specified.

In order to derive  equation for the determination
of a spectral index we  expand  
the phase function $R(\xi)$  in series of Legendre polynomials
(see also Sobolev 1975 and TL95)
$$
R(\xi)=p^0(\mu,\mu^{\prime})+2\sum_{m=1}^np^m(\mu,\mu^{\prime})
\cos{m(\varphi-\varphi^{\prime})},
\eqno(18)
$$
$$
p^m(\mu,\mu^{\prime})=\sum_{i=m}^nc_i^mP_i^{m}(\mu)P_i^{m}(\mu^{\prime})
\eqno(19)
$$
and 
$$
c_i^m=C_i{{(i-m)!}\over{(i+m)!}}
\eqno(20)
$$
for $m~=~0,~1,~2,~...~n$.
Since the phase function $R(\xi)$ is given by 
the series (Eq. 18) in $\cos{m\varphi}$,  
the source function  (the second term in brackets of the right hand side 
of Eq. 16), and $J (r,{\bf \Omega})$ can be 
expanded over $\cos{m\phi}$ too.
Under assumption of spherical symmetry for the source, we are interested in
the zero-term of the expansion which satisfies the following equations
$$
\ell J^0(r,\mu)=
-[n_e\sigma_T+(\alpha+3)\mu {{\partial\sqrt{f} }\over{\partial r}}]J^0(r,\mu)
+(n_e\sigma_T)B^0(r,\mu),
\eqno(21)
$$
where 
$$
\ell J^0(r,\mu)=\mu\sqrt{f}{{\partial J^0}\over{\partial r}}+
(1-\mu^2)\left({{\partial\sqrt{f} }\over{\partial r}}-
{{\sqrt{f} }\over{r}}\right){{\partial J^0}\over{\partial\mu}},
\eqno(22)
$$
and the source function 
$$
B^0(r,\mu)={1\over2}\int_{-1}^{1} p^0(\mu,\mu^{\prime})J^0(r,\mu^{\prime})
d\mu^{\prime}.
\eqno(23)
$$
There are two boundary conditions which our solution must satisfy.
The first is that there is no scattered radiation  outside of the 
atmosphere 
$$
J^0(0,\mu)=0~~~~~~~~{\rm for}~~~\mu<0.
\eqno(24a)
$$
The second boundary condition is that we have an absorptive boundary 
 at  radius $r_s$ 
$$
J^0(r_s,\mu)=0~~~~~~~~{\rm for}~~~\mu>0.
\eqno(24b)
$$

Thus the whole problem is reduced to the standard radiative transfer problem
for the space part of the solution $J(r,{\bf \Omega})$.
Inversion of the differential operator $\ell$ of the left hand side 
of equation (21) leads to the integral equation
for $B^0(r,\mu)$ 
$$
B^0(r,\mu)={1\over2}\int_{-1}^{0}p^0(\mu,\mu^{\prime})d\mu^{\prime}
$$
$$
\times \int_0^{T(r_{bn},r,\mu^{\prime})}
\exp\{-T[r_{bn},r^{\prime}(r,\mu^{\prime}),\mu^{\prime}]\}
B^0(r^{\prime},\mu^{\prime})dT 
$$

$$
+{1\over2}\int_{0}^{1}p^0(\mu,\mu^{\prime})d\mu^{\prime}
$$
$$
\times\int_0^{T(r_{bn},r,\mu^{\prime})}
\exp\{-T[r_{bn},r^{\prime}(r,\mu^{\prime}),\mu^{\prime}]\}
B^0(r^{\prime},\mu^{\prime})dT,
\eqno(25)
$$
where $T(r_{bn},r,\mu)$ is the optical path along characteristic curve of 
the differential operator $\hat\ell$ determined by the initial
point $r,~\mu^{\prime}$ toward the boundary radius $r_{bn}$ 
($r_{bn}=r_s$, and $\infty$ 
for the inner, and outer boundaries, respectively).

The phase function component $p^0(\mu,\mu^{\prime})$ entered in 
equations (21) and (23) is 
determined by the sum 
$$
p^0(\mu,\mu^{\prime})=\sum_{i=0}^nC_iP_i(\mu)P_i(\mu^{\prime}).
\eqno(26)
$$
Thus, we can present the source function $B^0$ also as a sum:
$$
B^0(r,\mu)=\sum_{i=0}^nC_iP_i(\mu)\int_{-1}^{1}P_i(\mu^{\prime})
J^0(r,\mu^{\prime})d\mu^{\prime}.
\eqno(27)
$$
This form of the source function is used for the solution of the
boundary problem (21-24) by the iteration method (e.g. Sunyaev \& Titarchuk 
1985). In order to proceed with the iteration method one has to assume 
some initial field distribution (in terms of the intensity $J^{0}$)
and then to calculate $B^0$ in accordance to Eq. (27) which 
is followed by the solution of the differential equation (21). 

This iteration formalism  is identical to the integral-equation formalism
in which    
$$
B^0(r,\mu)=\sum_{i=0}^nC_iP_i(\mu)B^0_i(r)
\eqno(28)
$$
where the set of $B^0_i(r)$, components of the source function 
$B(r,\Omega)$, is 
determined by the system of the integral equations (compare with TL95)
$$
B^0_i(r)= {1\over2}\sum_{j=0}^nC_j[\int_{-1}^{0}p_i(\mu^{\prime})
p_j(\mu^{\prime})d\mu^{\prime}
$$
$$
\times\int_0^{T(r_{bn},r,\mu^{\prime})}
\exp\{-T[r_{bn},r^{\prime}(r,\mu^{\prime}),\mu^{\prime}]\}
B^0_j(r^{\prime})dT 
$$
$$
+\int_{0}^{1}p_i(\mu^{\prime})p_j(\mu^{\prime})d\mu^{\prime}
$$
$$
\times\int_0^{T(r_{bn},r,\mu{\prime})}
\exp\{-T[r_{bn},r^{\prime}(r,\mu{\prime}),\mu^{\prime}]\}
B^0_j(r^{\prime})dT].
\eqno(29)
$$
 Thus the eigenvalue problem Eqs. (21-24), can be reduced 
to a eigenproblem for  a system
of  integral equations (29) where  
the optical paths $T(r_{bn},r,\mu)$ and the expansion coefficients of the 
phase function $C_i$ depend on the spectral index $\alpha$ as a parameter.
In other words, one has to find the values of $\alpha$  which guarantee
the existence of the nontrivial solution of equations (29).
In \S 3.2 and Appendix B  we shall proceed with the numerical solution of the 
eigenvalue problem by presenting 
 the bulk motion phase function $R_b(\xi_b)$ in the degenerated form 
(cf.  Eq. 26). 
\par
Now it is worth noting
that in the case of the pure thermal motion in the isothermal  
plasma cloud the problem is substantially simplified. 
 The source function  $B^0(r,\mu)$ 
can be replaced by its zeroth moment $B^0_0(r)$ (TL95) 
which guarantees the accuracy of the spectral index determination better 
than 10\% in the worst cases (Giesler \& Kirk 1997).

For example, the equation for the zeroth moment $B_0^0(r)$  reads  
$$
B^0_0(r)={{C_0}\over{2}}[\int_{-1}^{0}d\mu^{\prime}
\int_0^{T(r_{bn},r,\mu^{\prime})}
\exp\{-T[r_{bn},r^{\prime}(r,\mu^{\prime}),\mu^{\prime}]\}
B^0_0(r^{\prime})dT 
$$
$$
+\int_{0}^{1}d\mu^{\prime}
\int_0^{T(r_{bn},r,\mu{\prime})}
\exp\{-T[r_{bn},r^{\prime}(r,\mu{\prime}),\mu^{\prime}]\}
B^0_0(r^{\prime})dT].
\eqno(30)
$$
where $C_0$ is the zero-moment of the phase function.

\subsection[THE METHOD OF SOLUTION]{Photon trajectories and 
the  characteristics of the space operator $\ell$}

The characteristics of the differential 
operator $\ell$ are determined by the following differential 
equation
$$
\left[-{{1}\over{2x^2(1-x^{-1})}}+x^{-1}\right]dx=d[\ln(1-\mu^2)^{-1/2}],
\eqno(31)
$$
where $x=r/r_s$ is a dimensionless radius.
The integral curves of this equation (the characteristic curves) are given
by
$$
{{x(1-\mu^2)^{1/2}}\over{(1-x^{-1})^{1/2}}}=
{{x_0(1-\mu_0^2)^{1/2}}\over{(1-x_0^{-1})^{1/2}}}=p,
\eqno(32)
$$
where $p$ is an impact parameter at  infinity.
$p$ can  also be determined at a given point in a characteristic
 by  the cosine of an angle between the  tangent to and the radius vector to 
the point and by  the given point position $x_0$.
\par
\noindent
In the the flat geometry, 
the characteristics are just straight lines  
$$
x(1-\mu^2)^{1/2}=p,
\eqno(33)
$$
where an impact parameter $p$ is the distance of a given point to the center.  

We can resolve equation (32) with respect of $\mu$ to get
$$
\mu={\pm}(1-p^2/y^2)
\eqno(34)
$$
where $y=x^{3/2}/(x-1)^{1/2}$.
The graph of $y$ as a function of $x$ is presented in Fig. 1 which allows 
to comprehend the possible range of radii for the given impact parameter $p$ 
through the inequality $p\leq y$.
For example, if $p\leq \sqrt{6.75}$, then the photon can escape from the inner 
boundary (the black hole horizon) toward the observer or vice versa 
all photons going 
toward the horizon having these   
impact parameters are gravitationally attracted by the black hole. 
However, if $p>\sqrt{6.75}$,
then the finite trajectories are possible with the radius range between 
$1\leq x\leq 1.5$, or the infinite trajectories with $p\leq y(x)$ ($x$ 
is always more than $1.5$). 

\subsection[THE METHOD OF SOLUTION]{Spectral index determination} 

We are assuming a free fall for the background flow where the bulk
velocity of the infalling plasma is given by $v(r)=c(r_s/r)^{1/2}$. 
In the kinetic equations (12, 16) the density  $n$   is measured in 
the local rest frame of the flow and it is
$n=\dot m(r_s/r)^{1/2}/(2r\sigma_T )$. Here $\dot m=\dot M/\dot M_E$, 
$\dot M$ is  mass accretion rate  and 
 $\dot M_E \equiv L_E/c^2=4\pi GMm_p/ \sigma_Tc~$ is 
the Eddington accretion rate. 
                            
For the cold converging inflow ($kT_e=0$ keV) the electron distribution 
is the delta-function $F(r,P_e) =\delta({\bf v}-{\bf v}_b)$ defined 
in the velocity phase space in the way that
$$
\int_0^{\pi}\sin{\theta}d\theta\int d^3{\bf v}F(r,P_e)=1.
$$
 In this case the phase function is
$$
R_b(\xi)={3\over4} {{1}\over{\gamma_b^2}}
\left({{D_{1b}}\over{D_b}}\right)^{\alpha+2}{{1}\over{D_{1b}}}
[1+(\xi_b^{\prime})^2],
\eqno(35)
$$ 
where subscript ``b'' is related with the bulk  velocity direction
 (the case of arbitrary temperature will be considered elsewhere). 
In the case of zero temperature the directions of incoming and 
outcoming photons  are related.
Our goal is to find the nontrivial solution $J^0(r,\mu)$ 
of this homogeneous problem
and the appropriate spectral index $\alpha$ for which this  solution exists.
This problem can be solved  by the iteration method 
which involves the integration of the differential equation (21) 
with the given boundary conditions (24) along
the characteristics (32) by using Runge-Kutta's method.

The integration starts from the internal or the outer boundary depending 
on the particular impact parameter $p$ (Eq. 32). In turn which is determined 
by the dimensionless radius $x$ ($x=r/r_s$) and the cosine $\mu$ 
of the angle between the photon direction and the radius vector 
at the given point $x$. 

If $\mu$ is positive at the $x$ then the photon trajectory 
(the characteristics) can  start at the inner boundary  
[ if $x<1.5$ and $p<x^{3/2}/(x-1)^{1/2}$ 
or if  $x>1.5$ and $p<\sqrt{6.75}$] or at the outer boundary 
(if  $x>1.5$ and $p>\sqrt{6.75}$).

%But if  $x\geq 1.5$ and $\sqrt{6.75}\leq p\leq x^{3/2}/(x-1)^{1/2}$
%the photon trajectories (characteristics ) start at  infinity. 

All cases can be understood from  Fig. 1. 
The trajectories with the given $p$ are related to the parallel  
lines to the X-axis, $y=p$. These lines  start at $x=1$ or at 
 infinity. For example if they start at  infinity (i.e. having 
negative $\mu$) and  $p\geq \sqrt{6.75}$ 
 they must have the turning point with $\mu=0$. Thus they
must pass through the point with radius $x_{\star}$ where 
 $p=x_{\star}^{3/2}/(x_{\star}-1)$. 
At this point the cosine $\mu$ 
changes sign from $-$ to $+$ and after that the trajectory enters through the 
point with radius $x$ at the positive angle 
$\theta=\cos^{-1}\mu$.  

\placefigure{fig1}

If the trajectory starts at $x=1$ , (i.e. having
positive $\mu$) and $p< \sqrt{6.75}$ 
the parallel line $y=p$ has no turning point. 

If $\mu$ is negative at $x$, the trajectories  starting
at the internal boundary (having positive cosine $\mu$)
have to pass through the turning point $\mu=0$ (changing the
cosine sign) at  radius $x_{\star}$ where  
$p=x_{\star}^{3/2}/(x_{\star}-1)$. However if the trajectory   
starts at the outer boundary has no the turning points the cosine $\mu$ is
always negative along the trajectory.   
   
This space integration is followed by integration $J^0(r,\mu^{\prime})$
 over the angular variable $\mu^{\prime}$ in Eq. (23).
As  the initial 
distribution for $B^0(r,\mu)$ or $J^0(r,\mu)$ 
we can choose, for example, the uniform  one.
We use the Gaussian integration to calculate of Eq. (23) 
(see e.g. Abramovitz \& Stegan 1970 for details of the methods). 
After quite a few iterations the iterative 
process converges and  it produces the eigenfunction source distribution
$B^0(r,\mu)$. The number of iterations $n$ is related to the average number
of scatterings which the soft photons undergo to transform into the hard 
ones (e.g. Titarchuk 1994). It is determined by the 
Thomson optical depth of the bulk motion atmosphere $\tau_b=\tau_T(r_s)$.
 For the cold atmosphere ($T_e=0$) 
the iteration number $n\simeq 2\tau_b$.  The convergence of the process 
can be done only with the proper choice of the value of a spectral index 
$\alpha$.

\section{ RESULTS OF CALCULATIONS AND DISCUSSION}

 Fig. 2 presents the results of
the calculations of the  spectral  indices as a function
of mass accretion rates. 
It is clearly seen 
that the spectral index is a weak function of  mass accretion rate
in a wide range of $\dot m =3-10$. The asymptotic value of the spectral
index for high mass accretion rate is 1.75 which is between $\alpha =2$ 
that was found by Blandford \& Payne 1981 for the infinite medium and
$\alpha\approx 1.4$ which was found by TMK96 for the finite bulk motion 
atmosphere. 

\placefigure{fig2}

The latter two results are obtained in the nonrelativistic Fokker-Planck 
approximation.
We see that the efficiency of the hard photon production  in the cold 
bulk motion atmosphere is quite low. This is not the case if the plasma 
temperature is of order of a few keV or higher. The coupling effect 
between the bulk  and local Maxwellian motion occurs when the bulk
motion velocity is very close to the speed of light, i.e. when 
the matter is very close to the horizon. 
The upscattering effect increases significantly in the latter case. 
 In the regime of the relativistic bulk motion 
the  electron distribution (2) has a sharp maximum at $\theta=\pi$ and $V=c$.
In the vicinity of the maximum the  distribution is characterized by 
the exponential shape $F(r. P_e)\propto \exp(-\beta \gamma_b/2\gamma)$
(see also Eq. 5).  
More results and details regarding the relativistic coupling would be 
presented elsewhere.

\placefigure{fig3}

As an example, in Fig. 3 we demonstrate  the zeroth  moment of the 
source function distribution (the hard photon production). It is seen there 
that the distribution  has a strong peak around 
2 $r_s$. This  means that the vicinity of the black hole is  a place 
where the  hard photons are produced by upscattering of the soft photons 
off the converging electrons. 

Our calculations were made under the assumption of the free fall velocity 
profile.
Since the energy gain due to the bulk motion Comptonization is not bigger 
than factor 3 (if the spectral indices are higher than 1.5, TMK97), it 
follows that we can safely neglect the effects of the radiation force in
our calculations if the injected photon flux in the converging inflow 
is of order of a few percent of the Eddington luminosity.

The assumption of Thomson scattering  accepted in our solution restricts 
the relevant energy range to $E<m_ec^2$. 
Our approach cannot determine accurately the exact position of the high 
energy cutoff which is formed due the downscattering of the very energetic 
photons in the bulk motion electron atmosphere. The additional efforts 
are required to confirm the qualitative estimates of the high energy cutoff
position as of order $m_ec^2$ (TMK97). Laurent \& Titarchuk 1997 (in 
preparation) by using Monte Carlo calculations checked and confirmed our 
results for the spectral indices and the TMK97 
estimates of the high energy cutoff position. 
Futhermore,  they found the prominent spectral features at 
energies $\gtorder 400$ keV.  

As a conclusion  we would like to point out the definitive 
(according to our model)  difference  between black holes and neutron
stars, as it can be ascertained in their spectral properties while
in their soft  states, when their luminosity is dominated
by the quasithermal, soft, component: In the black hole case  there 
should always be an additional steep power law high energy tail extending
to energies $\sim m_e c^2$. This component should 
be absent in neutron star systems, because the effect of the bulk 
motion is suppressed by the radiation pressure in this case.

We presented the full relativistic formalism and solved 
semi-analytically the kinetic equation by using TL95 eigenfunction method 
(see also Giesler \& Kirk 1997)  in the case of  plasma 
infalling radially into a compact object with a soft source of input photons. 
We found  that {\it the converging 
flow has crucial effects on the emergent spectrum for moderately 
super-Eddington mass accretion rates}. 

Our power law spectra can be applicable for the explanation of 
 the observational situations in black hole candidate sources.

\section{ACKNOWLEDGMENTS}

We thank the anonymous referee for reading and evaluating the present
paper. 
L.T. would like to acknowledge support from,  NASA grants NCC5-52, NAG 5-
3408 and Alex Muslimov and  Leonid Ozernoy for discussions and 
useful suggestions. L.T. also acknowledges Wan Chen for pointing out 
some particular details of the observational situations in black hole 
candidate sources.

\newpage

\appendix

\section{GENERAL RELATIVISTIC RADIATIVE KINETIC FORMALISM} 

Let ${\bf u}$ be an arbitrary time like unit vector (${\bf u\cdot u}
\equiv u^{\alpha}u_{\alpha}=-1$, $u^0>0$) at some given space time 
point $x$, and let ${\bf d}_1x,{\bf d}_2x,{\bf d}_3x$ be the three 
arbitrary displacement vectors (with components $d_1x^{\alpha}$, etc) 
which span an element of hypersurface orthogonal to ${\bf u}$. 
By using orthogonality of two vectors  $ u_{\lambda}$ and 
$${d{\Sigma}}_{\lambda}=
\sqrt{(-g)}\epsilon_{\alpha\beta\gamma\lambda}d_1x^{\alpha}
 d_2x^{\beta}d_3x^{\gamma}
\eqno(A1)
$$
 to the  element of the hypersurface one can 
get the invariant volume element orthogonal to the unit vector ${\bf u}$
as follows:
$$
dV=\sqrt{(-g)}\epsilon_{\lambda\alpha\beta\gamma}u^{\lambda}d_1x^{\alpha}
 d_2x^{\beta}d_3x^{\gamma},
\eqno(A2)
$$
where $ g={\rm det}|g_{\alpha\beta}|$ , $g_{\alpha\beta}$ is the metric tenzor
(see Eq. 1) and $\epsilon_{\lambda\alpha\beta\gamma}$ is the Levi-Civita
alternating symbol  $\epsilon_{0123}=+1$.
Since $N$ is independent of $\bf u$ we may orient $dV$ to make $u^{0}=1, 
~u^{i}=0$. Thus in a local Minkowski frame it would be
$$
 dV=d^3 x.
\eqno(A3)
$$
Similarly, at the point ${\bf p}$ (${\bf p}\cdot{\bf p}=0$) 
the 3-surface element of the zero-mass shell 
  spanned by three 
displacement vectors  ${\bf d}_1p,{\bf d}_2p,{\bf d}_3p$
is presented by
$$
(dP) p_{\lambda}=
\sqrt{(-g)}\epsilon_{\alpha\beta\gamma\lambda}d_1p^{\alpha}
 d_2p^{\beta}d_3p^{\gamma}.
\eqno(A4)
$$
Thus the invariant volume $dP$ is 
$$
dP = \sqrt{(-g)}\epsilon_{ijk}{{d_1p^{i}d_2p^{j}d_3p^{k}}\over{-p_0}}.
\eqno(A5)
$$
because all products of the right hand side of equation (A4) with $p^{\lambda}$ 
for $\lambda\not=0$ are equal to zero [this follows from 
equality  ${\bf p\cdot\bf p}=0$, i.e. $dp^0=
(g_{ij}p^{j}dp^i)/g_{0\alpha}p^{\alpha 0}$][B
 and the fact that
a numbers of the even and odd transpositions of the combination 
$\alpha\beta\gamma$ are equal). 
When local Minskowian coordinates are used, 
with ${\bf d}_1p,{\bf d}_2p,{\bf d}_3p$
tangent to the respective coordinate lines, this  reduces to the familiar 
Lorentz invariant $3-$surface element in the momentum space, $d^3p/E$ 
(e.g. Landau \& Lifshitz 1971). If we introduce in the zero-shell spherical 
coordinates $(E,~\theta, \varphi)$ defined (in a local Minkowski frame)
through relationships
$$
p^1=E\cos\theta,~~~~~p^2=E\sin\theta\cos\varphi,~~~~~
p^3=E\sin\theta\sin\varphi
\eqno(A6)
$$
where $E$ is a photon energy, then we obtain 
$$
dP=pdEd\Omega. 
\eqno(A7)
$$
In any coordinate system one can introduce the invariant infinitesimal 
volume element over the photon space, erected above
 the space time point under consideration. In other words 
one can introduce
 an orthonormal tetrad in each point of space-time  and get the 
same result as equation (A7).
 
The number of photon world lines crossing 
an infinitesimal ``3-area'' $d{\Sigma}_{\lambda}$ (or $dV$)
 at $x$, with 4-momenta in the range $dP$ is 
%$$
%(P^{a}, d{\omega})=
%(P^{a},{(-g)^{1\over 2} \over |p_{0}|}dp^{1}dp^{2}dp^{3})\eqno(2)$$
%that cross 
%in the spacetime and in the direction of its normal $n_{\mu}$ is given by:

$$
dN=N(x,{\bf p})=N(x,{\bf p})(-{\bf p\cdot u})dPdV
\eqno(A8)
$$
where the inclusion of the projector $(-{\bf p\cdot u})$ 
takes care of perpendicular counting of photon states along 
${d{\Sigma}}_{\lambda}$ or $dV$.

%The quantity
%$$d{\omega}={(-g)^{1\over 2} \over |p_{0}|}dp^{1}dp^{2}dp^{3})\eqno(4)$$
%is
 
The above somewhat abstract but fully covariant definition of the 
photon distribution function $N$  incorporates, in fact, the familiar density
interpretation of the classical Boltzmann distribution function
defined over the six-dimension classical phase space of 
the photon gas. One simply has to take  $dV$ along an arbitrary 
 $t=const$ (see Eq. A3) spacelike hypersurface equipped with orthogonal 
coordinates, to recover from (A8) and (A7)  that
$$
dN=N(x,{\bf p})d^{3}xd^{3}P,
\eqno(A9)
$$
i.e. the familiar phase space density interpretation of 
the distribution function. Comparing (A9) with the classical definition, 
$$
dN=f(x,E,\Omega)dE~d\Omega d^3x,
\eqno(A10)
$$
($\Omega$ is the unit 3-vector in the direction of the beam)
one can obtain that
$$
N(x,{\bf p})=E^{-2}f(x,E,\Omega)
\eqno(A11)
$$
in the local proper frame. Since $N$ is a scalar, this gives its value 
in any other frame as well; it can be used to infer the transformation
properties of $f$ and the specific intensity $I_{\nu}=E f=E^3 N$.

The general relativistic transfer equation essentially describes 
the rate 
of change of the above defined number of world lines $dN$
as it is pushed  along an 
arbitrary null spacetime geodesic.
In the presence of  external fields, the number of photon world lines
 centered around $\bf p$, is altered. The 
alteration depends upon the specific interaction the
photon field is subjected to. In accordance with the relativistic form of 
Liouville's Theorem (e.g. Lindquist 1966)
${(-\bf p\cdot u)}dV dP$ remains invariant along 
the given set of trajectories. 
Hence the change in the number of world lines within $({-\bf p\cdot u})dV 
dPd\tau=dW~dP$, where $\tau$ is some parameter changing along geodesics 
(Landau \& Lifshitz 1971), is
$$
{{dx^{\alpha}}\over{d\tau}}=p^{\alpha},~~~~
{{dp^{\alpha}}\over{d\tau}}=-\Gamma^{\alpha}_{\beta\gamma}p^{\beta}p^{\gamma}
\eqno(A12)
$$
which is simply  proportional to the change in $N$.
Thus  the  transfer equation takes the following form 
$$
p^{\alpha}{{DN}\over{dx^{\alpha}}}~=~
p^{\alpha}{{\partial N}\over{\partial x^{\alpha}}}-
{{\partial N}\over{\partial p^{\beta}}}
\Gamma^{\beta}_{\alpha\gamma}p^{\alpha}p^{\gamma}~=~S(N)
\eqno(A13)
$$
where a product $S(N)~dW~dP$ denotes  the change dN in  all possible 
interactions 
taking place between the radiation field and external sources within 
$dW$.   $\Gamma^{\alpha}_{\beta\gamma}$ are the Christoffel symbols.

In the present paper we shall be interested exclusively in 
the Compton scattering of the radiation field of a locally Maxwellian
distribution of electrons representing the plasma inflowing into black hole.
 To deal with the general covariant of the theory we shall 
perform all the calculations relative to the local orthonormal frame or
tetrad $\{{\bf e}_a(x)\}$ ($a=0,1,2,3$) 
tied to the metric (1) (e.g. Landau \& Lifshitz 1971). 
Relative to
such a local orthonormal frame,
an arbitrary photon momentum ${\bf p}$ may be represented as
$$
{\bf p}=p^{a}{\bf e}_a.
\eqno(A14)
$$   
If $p^{\alpha}$ is the usual contravariant components of ${\bf p}$ in some
coordinate system $\{x^{\alpha}\}$ then 
$$
{\bf p}=p^{\alpha}{\bf e}_{\alpha}.
\eqno(A15)
$$   
where $\{{\bf e}_{\alpha}\}$ is the induced coordinate basis.
If we define
$$
{\bf e}_a=e_a^{\alpha}{\bf e}_{\alpha}~~~~{\rm and}~~~~
{\bf e}_{\alpha}=e^a_{\alpha}{\bf e}_{a},
\eqno(A16)
$$   
then it is evident that
$$
p^a=e^a_{\alpha} p^{\alpha}~~~~{\rm and}~~~~
 p^{\alpha}=e_a^{\alpha} p^{a}.
\eqno(A17)
$$
correspondingly.
One can introduce the transformation which leaves the coordinate 
system unaltered, but express vectors in terms of their tetrad components,
namely
$$
x^{\prime\alpha}=x^{\alpha},
$$
$$
p^{\prime a}=e^a_{\alpha} p^{\alpha}.
\eqno(A18)
$$
By using this transformation 
the radiative transfer equation can be rewritten in the form,
which is very similar to Eq. (A13)
$$
p^{a}e_a^{\alpha}{{\partial N}\over{\partial x^{\alpha}}}-
{{\partial N}\over{\partial p^{b}}}
\gamma^{b}_{ac}p^{c}p^{a}~=~S(N),
\eqno(A19)
$$
with 
$$
\gamma^{b}_{ac}= e_a^{\alpha}e_{\gamma}^{b}e_{a;\alpha}^{\gamma}
\eqno(A20)
$$
being the Ricci rotation coefficients (e.g. Landau \& Lifshitz 1971).

We introduce in the tangent space, at each event $(t,r,\theta,\varphi)$, the 
orthormal basis
$$
{\bf e}_0=e^{\Lambda/2}{\bf e}_t,~~~~{\bf e}_1=e^{-\Lambda/2}{\bf e}_r~~~~
{\bf e}_2=r^{-1}{\bf e}_{\theta}~~~~{\bf e}_3=(r\sin\theta)^{-1}
{\bf e}_{\varphi},
\eqno(A21)
$$
where $e^{\Lambda/2}=f$ [see eq. (1)].
Relative to this local orthormal frame an arbitrary photon momentum can be 
represented as 
$$  
{\bf p}=[p^{(0)},p^{(1)},p^{(2)}
p^{(3)}]=(E,E\cos{\theta},E\sin{\theta}\cos{\varphi},
E\sin{\theta}\sin{\varphi}) 
\eqno(A22)
$$
where $E$ is the physical photon energy as measured by an
 orthonormal observer, and ${\theta}, {\varphi}$ are local polar-like
phase space coordinates. The angle ${\theta}$ is measured
relative to the outward radial direction at the point under consideration. 
For our problem we shall assume the steady state conditions
for the accretion and radiation field, respectively, implying that
 $N$ is time-independent and spherically symmetric. Furthermore, 
$N$ should be axially symmetric  around the radial direction, 
which results in  $N=N(r,{E}, {\mu})$ where ${\mu}=cos{\theta}$.

Thus the coefficients $e_a^{\alpha}$ and $e^a_{\alpha}$ are given by
$$
e_1^{\alpha}=\delta_1^{\alpha}e^{\Lambda/2},~~~~
e_2^{\alpha}=\delta_2^{\alpha}e^{-\Lambda/2},
$$       
$$
e_3^{\alpha}=\delta_3^{\alpha}/r,~~~~
e_4^{\alpha}=\delta_2^{\alpha}/(r\sin\theta),
\eqno(A23)
$$       
and  
$$
e_1^{a}=\delta_1^{a}e^{-\Lambda/2},~~~~
e_2^{a}=\delta_2^{a}e^{\Lambda/2}.
$$       
$$
e_3^{a}=\delta_3^{a}r,~~~~
e_4^{a}=\delta_4^{a}r\sin\theta.
\eqno(A24)
$$       
Here $\delta_i^j$ is Kronecker delta ($=0$ if $i\not=j$; 
$=1$ if $i=j$).

By using these formulas one can calculate the rotational coefficients
$\gamma^{b}_{ac}$ [see eq. (A20)] and get the relativistic radiative 
transfer equation in the form 
$$
e^{-\Lambda/2}\left [\mu {{\partial N}\over{\partial r}}
+{{(1-\mu^2)}\over{r}} {{\partial N}\over{\partial \mu}}\right]
+e^{-\Lambda/2}{{\Lambda^{\prime}}\over {2}} 
\left[\mu E{{\partial N}\over{\partial E}}
+(1-\mu^2){{\partial N}\over{\partial \mu}}\right]
={{S(N)}\over{E}}.
\eqno(A25)
$$
The  terms in the first bracket are obtained by using formulas (A24) 
for $e_{(1a)}^{(1\alpha)}$ and $e_{(2a)}^{(2\alpha)}$ and the fact that
the derivative 
$$
{{d\mu}\over{d\theta}}= -\sin\theta e^{-\Lambda/2}.
$$
The first term in the second bracket is related with the Ricci coefficient
$\gamma_{00}^{1}$ and the second term there appears as a result of 
calculation of product $\gamma_{00}^{1}(p^{0})^2(\partial N/\partial p^1)$.
We also use the fact that the partial derivative 
$$
{{\partial N}\over{\partial p^1}}={{1-\mu^2}\over {E}}
{{\partial N}\over{\partial \mu}}+\mu{{\partial N}\over{\partial E}}.
\eqno(A26)
$$ 

\clearpage

\section{NUMERICAL CALCULATION OF THE SOURCE FUNCTION} 

We use the iteration method to solve the radiative transfer problem 
Eqs. (21-24) which involves the calculation of the source function $B^0$. 
In order to calculate the source function we transform it into 
the degenerate form by using the phase function formula (35).
We have 
$$
B^0(r,\mu)={1\over2}\int_{-1}^{1} p^0(\mu,\mu^{\prime})J^0(r,\mu^{\prime})
d\mu^{\prime}=(I_1+I_2 +I_4+I_7) + (I_3+I_5)\mu +(I_6-I_7)\mu^2.
\eqno( B1)
$$
The following integrals $I_i$ are computed by using the Gaussian 
integration formula (see for details of the Gaussian integration method, e.g.
Stegan \& Abramovitz 1966).
$$
I_1={3\over4} {{1}\over{\gamma_b^2}}
{{1}\over{D_b^{\alpha+2}}}\int_{-1}^{1}D_{1b}^{\alpha+1}
J^0(r,\mu^{\prime})d\mu^{\prime}
\eqno(B2)
$$
$$
I_2=-{3\over4} {{1}\over{\gamma_b^4}}
{{1}\over{D_b^{\alpha+3}}}\int_{-1}^{1}D_{1b}^{\alpha}
J^0(r,\mu^{\prime})d\mu^{\prime}
\eqno(B3)
$$
$$
I_3={3\over4} {{1}\over{\gamma_b^4}}
{{1}\over{D_b^{\alpha+3}}}\int_{-1}^{1}\mu^{\prime}D_{1b}^{\alpha}
J^0(r,\mu^{\prime})d\mu^{\prime}
\eqno(B4)
$$
$$
I_4={3\over8} {{1}\over{\gamma_b^6}}
{{1}\over{D_b^{\alpha+4}}}\int_{-1}^{1}D_{1b}^{\alpha-1}
J^0(r,\mu^{\prime})d\mu^{\prime}
\eqno(B5)
$$
$$
I_5=-{3\over4} {{1}\over{\gamma_b^6}}
{{1}\over{D_b^{\alpha+4}}}\int_{-1}^{1}\mu^{\prime}D_{1b}^{\alpha-1}
J^0(r,\mu^{\prime})d\mu^{\prime}
\eqno(B6)
$$
$$
I_6={3\over8} {{1}\over{\gamma_b^6}}
{{1}\over{D_b^{\alpha+4}}}\int_{-1}^{1}(\mu^{\prime})^2D_{1b}^{\alpha-1}
J^0(r,\mu^{\prime})d\mu^{\prime}
\eqno(B7)
$$
$$
I_7={3\over16} {{1}\over{\gamma_b^6}}
{{1}\over{D_b^{\alpha+4}}}\int_{-1}^{1}[1-(\mu^{\prime})^2]D_{1b}^{\alpha-1}
J^0(r,\mu^{\prime})d\mu^{\prime}
\eqno(B8)
$$

\newpage

\figcaption[fig1.eps]
{Plot of the photon trajectory phase space. The phase space 
function y versus the dimensionless radius $x=r/r_s$. \label{fig1}}

\figcaption[fig2.eps]
{Plot of the  energy spectral index
(photo index minus one) versus  the total mass accretion rate.
\label{fig2}}
 
\figcaption[fig3.eps]
{Plot of the source function distribution, Eq. (B1) in 
arbitrary units versus the dimensionless radius $x=r/r_s$  
for the dimensionless mass accretion rate $\dot m=4$ and $\mu=0$. 
\label{fig3}}

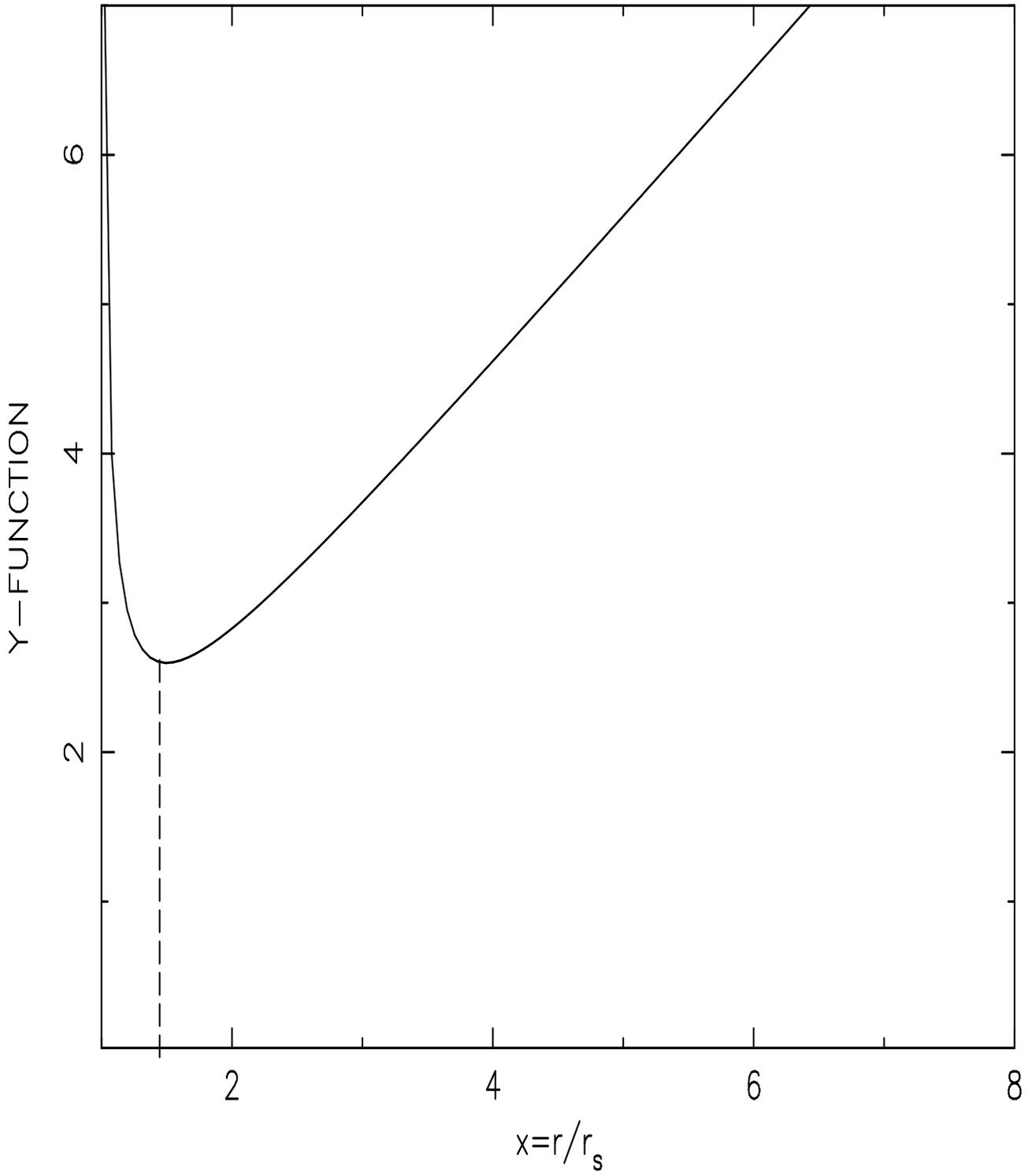
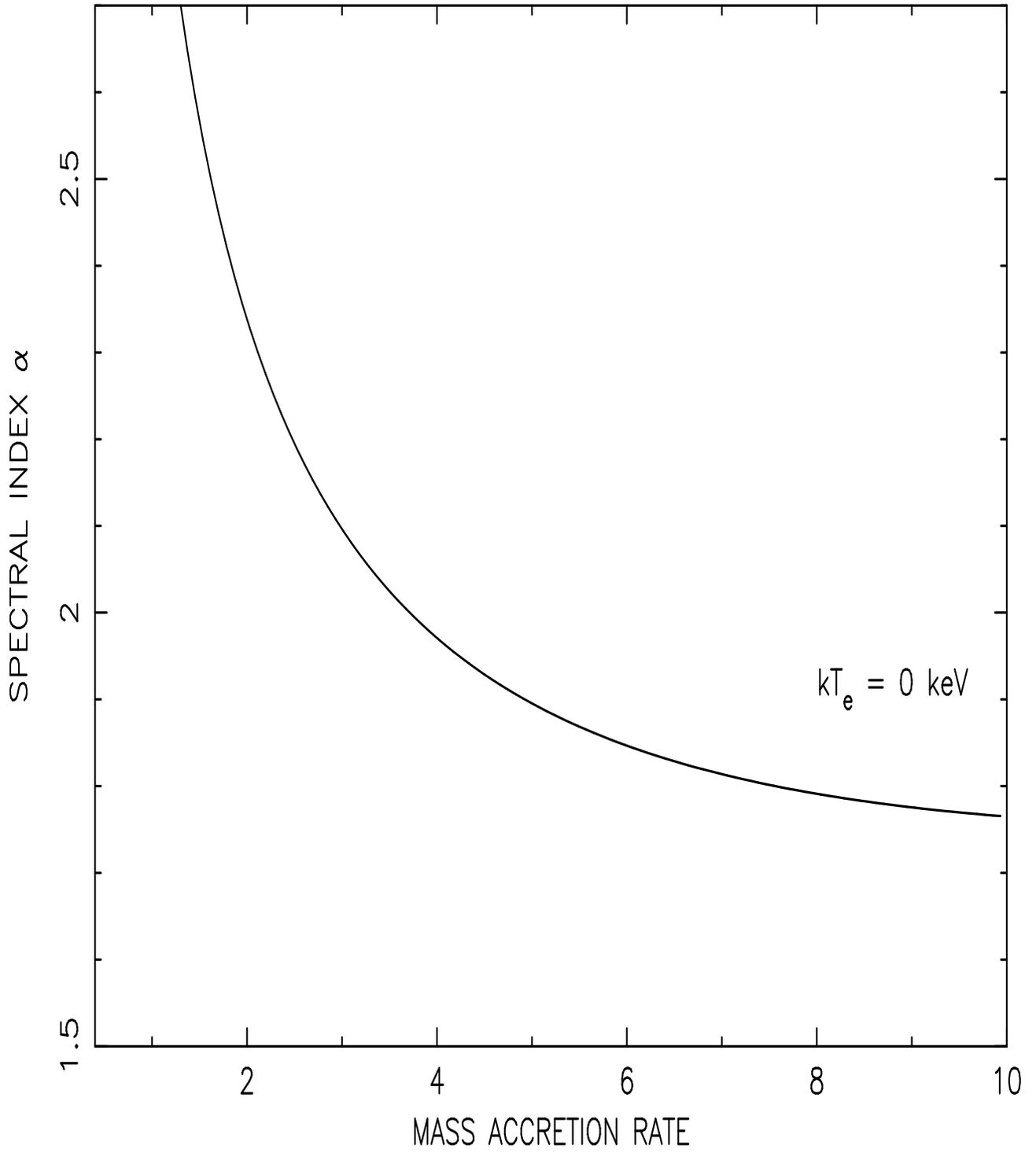
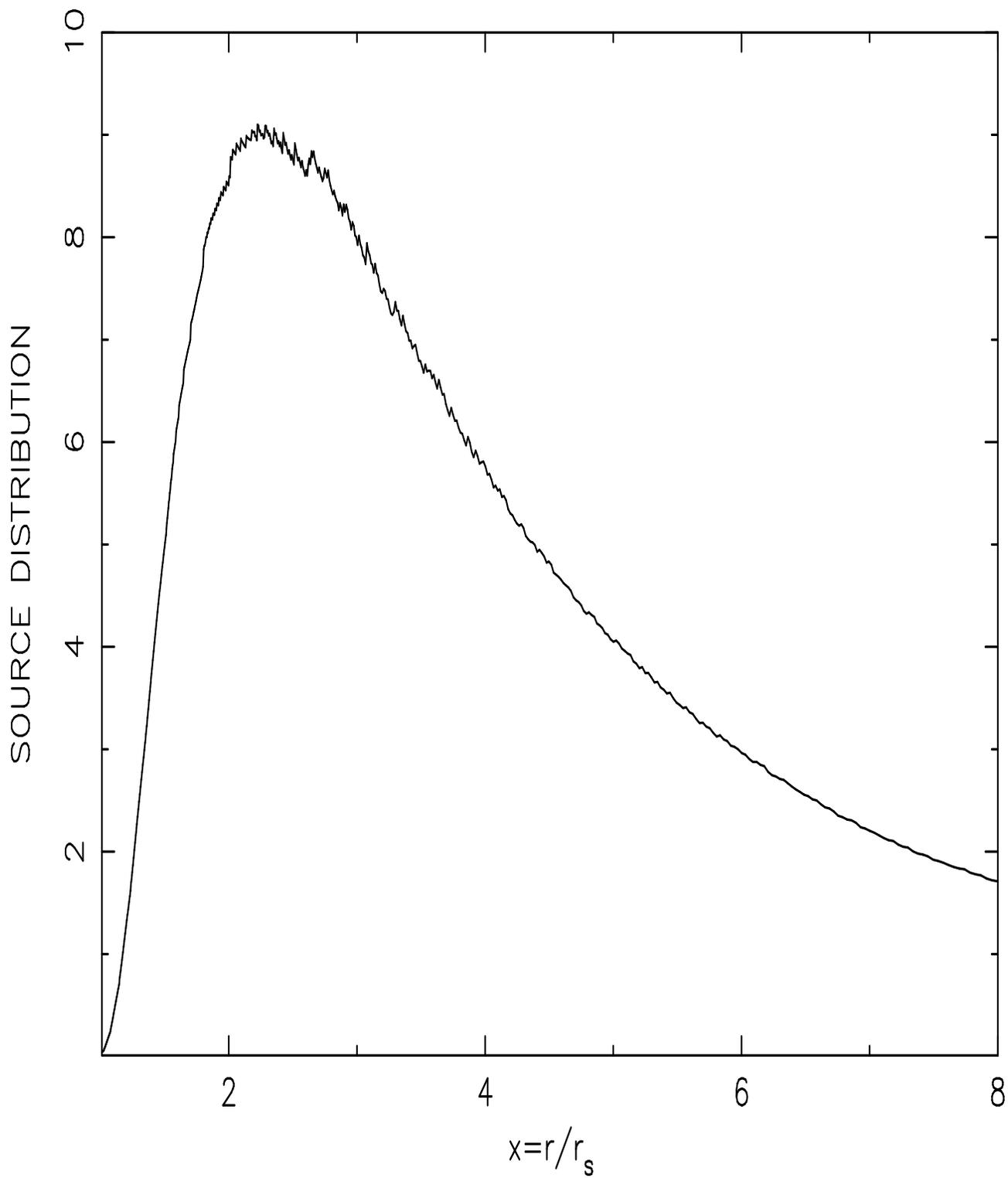
\end{document}